\documentclass[reprint,
noshowpacs,
 superscriptaddress,
 amsmath,amssymb,
 floatfix,
 twocolumns
]{revtex4-1}
\usepackage[utf8]{inputenc}
\usepackage[T1]{fontenc}

\usepackage{float}
\usepackage{physics}
\usepackage{graphicx}\usepackage{dcolumn}\usepackage{bm}\usepackage[x11names]{xcolor}
\usepackage{booktabs}
\usepackage{dsfont}
\usepackage{mathrsfs}
\usepackage{amssymb}
\usepackage[normalem]{ulem}
\usepackage{csquotes}

\usepackage{hyperref}
\usepackage{bookmark}

\definecolor{erk_pink}{HTML}{DB1C8E}

\hypersetup{
  breaklinks=true,
  colorlinks=true,
  linkcolor=erk_pink,
  filecolor=black,
  urlcolor=erk_pink,
  citecolor=erk_pink,
  pdfborder=0 0 0,
}

\newcommand{\height}{10pt}
\newcommand{\cell}{\raisebox{-0.25\height}{\includegraphics[height=\height]{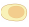}}}
\newcommand{\state}{\raisebox{-0.25\height}{\includegraphics[height=\height]{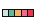}}}
\newcommand{\drift}{\raisebox{-0.25\height}{\includegraphics[height=\height]{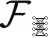}}}
\newcommand{\noise}{\raisebox{-0.25\height}{\includegraphics[height=\height]{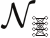}}}
\newcommand{\seq}{\raisebox{-0.25\height}{\includegraphics[height=\height]{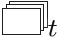}}}
\newcommand{\graph}{\raisebox{-0.25\height}{\includegraphics[height=\height]{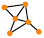}}}
\newcommand{\prob}{\raisebox{-0.25\height}{\includegraphics[height=\height]{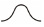}}}

\newcommand{\G}{\mathcal{G}}
\newcommand{\V}{\mathcal{V}}
\newcommand{\E}{\mathcal{E}}
\newcommand{\N}{\mathcal{N}}

\newcommand{\x}{\mathbf{x}}
\newcommand{\h}{\mathbf{h}}
\newcommand{\m}{\mathbf{m}}
\newcommand{\z}{\mathbf{z}}

\newcommand{\comp}{\text{c}}
\newcommand{\mess}{\text{m}}

\newcommand{\Gc}{\G^\comp}
\newcommand{\Vc}{\V^\comp}
\newcommand{\Ec}{\E^\comp}

\newcommand{\Gm}{\G^\mess}
\newcommand{\Vm}{\V^\mess}
\newcommand{\Em}{\E^\mess}

\makeatletter\begin{document}

\title{Learning tissue dynamics with multiscale stochastic graph neural networks}
\title{Learning noisy tissue dynamics across time scales}

\author{{Ming Han}}
\affiliation{Center for Quantitative Biology and Peking-Tsinghua Joint Center for Life Sciences, Academy for Advanced Interdisciplinary Studies, Peking University, Beijing, China}
\author{{John Devany}}
\affiliation{James Franck Institute and Department of Physics, The University of Chicago, Chicago, Illinois 60637, USA}
\affiliation{Feinberg School of Medicine, Northwestern University, Chicago, IL 60611, USA} 
\author{{Michel Fruchart}}
\affiliation{Gulliver, ESPCI Paris, Université PSL, CNRS, 75005 Paris, France}
\author{{Margaret L. Gardel}}
\affiliation{James Franck Institute and Department of Physics, The University of Chicago, Chicago, Illinois 60637, USA}
\affiliation{Molecular Genetics and Cell Biology, The University of Chicago, Chicago, Illinois, 60637, USA}
\affiliation{Pritzker School of Molecular Engineering, The University of Chicago, Chicago, Illinois, 60637, USA}
\affiliation{Chan Zuckerberg Biohub Chicago, Chicago, IL, USA}
\author{Vincenzo Vitelli}
\affiliation{James Franck Institute and Department of Physics, The University of Chicago, Chicago, Illinois 60637, USA}
\affiliation{Leinweber Institute for Theoretical Physics, University of Chicago, Chicago, Illinois 60637, USA}.
\affiliation{Chan Zuckerberg Biohub Chicago, Chicago, IL, USA}

\begin{abstract}
Tissue dynamics play a crucial role in biological processes ranging from inflammation to morphogenesis. 
However, these noisy multicellular dynamics are notoriously hard to predict.
Here, we introduce a biomimetic machine learning framework capable of inferring noisy multicellular dynamics directly from experimental movies. 
This generative model combines graph neural networks, normalizing flows and WaveNet algorithms to represent tissues as neural stochastic differential equations where cells are edges of an evolving graph. Cell interactions are encoded in a dual signaling  graph capable of handling signaling cascades.
The dual graph architecture of our neural networks reflects the architecture of the underlying biological tissues, substantially reducing the amount of data needed for training, compared to convolutional or fully-connected neural networks.
Taking epithelial tissue experiments as a case study, we show that our model not only captures stochastic cell motion but also predicts the evolution of cell states in their division cycle. 
Finally, we demonstrate that our method can accurately generate the experimental dynamics of developmental systems, such as the fly wing, and cell signaling processes mediated by stochastic ERK waves, paving the way for its use as a digital twin in bioengineering and clinical contexts. \end{abstract}

\maketitle

The physicist Eugene Wigner famously commented on the \enquote{unreasonable effectiveness of mathematics in the natural sciences} \cite{Wigner1960}.
Continuum theories are an example of such unreasonable effectiveness~\cite{Gaspard2022}. These deterministic, memoryless, coarse-grained theories describe, often with uncanny precision, systems that are discrete, stochastic, and non-Markovian. Yet, this triad of complexity comes back to haunt us when we seek top-down approaches that infer microscopic rules, including single-cell variability and noise, from the macroscopic behavior of biological systems. Inspired by Wigner's creed, here we ask: What modeling framework, if any, can reliably predict the dynamics of noisy multicellular systems across time scales?

\begin{figure*}[b]
\includegraphics[width=0.95\textwidth]{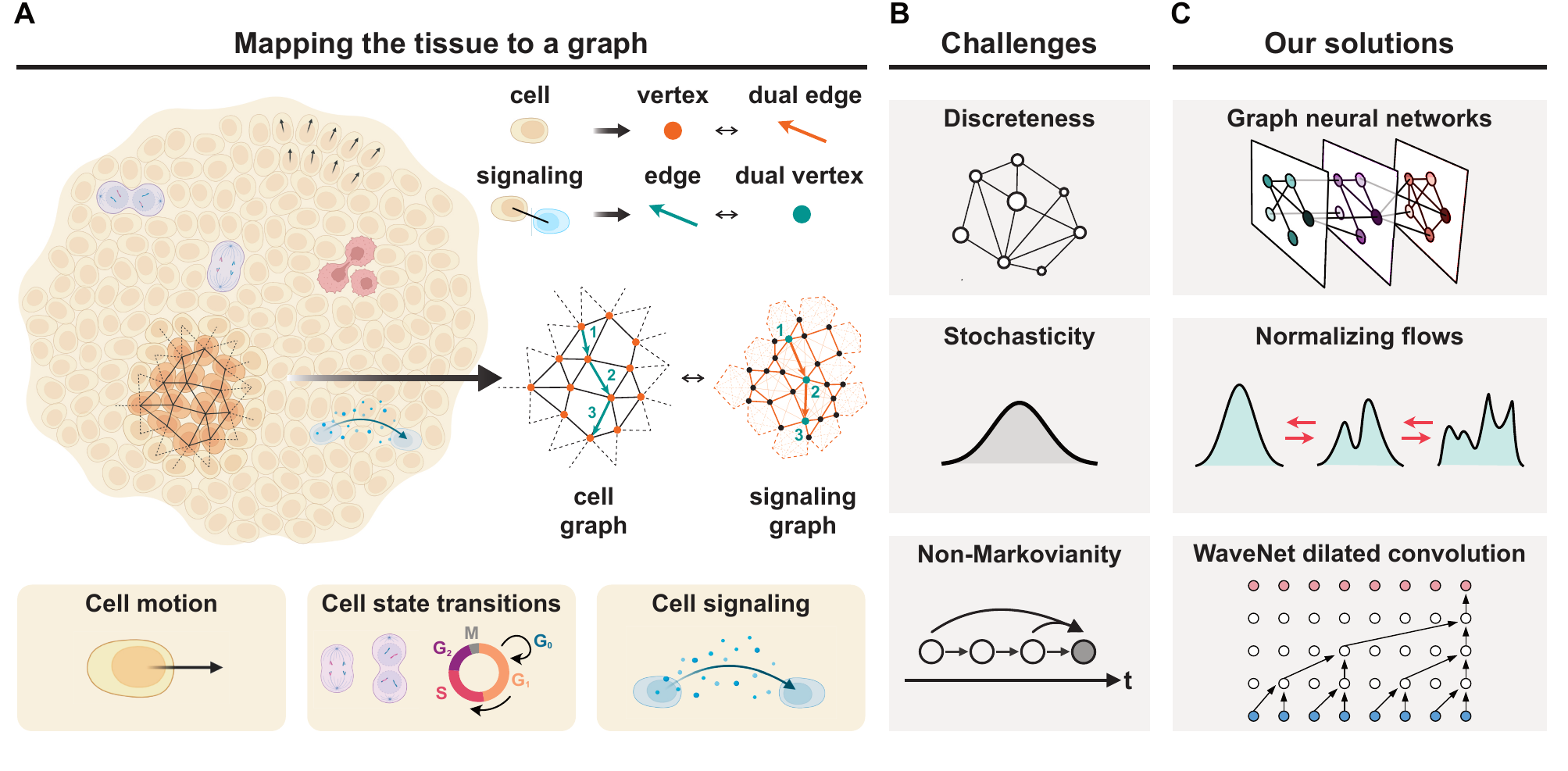}
\caption{\label{fig_gwn} \textbf{Deep learning for noisy multicellular dynamics.} 
\textbf{A.}
Cells in a tissue undergo three characteristic dynamical behaviors, motion, cell state transitions, and cell signaling (bottom row). 
These cellular dynamics are often subject to intrinsic stochasticity due to the complex physical and chemical processes occurring inside the cells.
As the basic structural and functional unit of living organisms, cells also exhibit strong discreteness in many aspects, ranging from their granularity in space to cell division or state change in time. 
Moreover, through stress response and chemical signaling, cells experience complex interactions from their neighbors, which could even in turn regulate the internal states of the cells. 
The cell network in a tissue is naturally represented as a cell graph where the nodes are the nuclei of the cells and the edges denote their interconnectivity through touching membranes.
In practice, it is also convenient to construct a signaling graph with structure dual to the cell graph, which is used in the graph neural networks to describe interactions (see Methods for details). 
In the dual graphs, the vertices and edges are interchanged, so the information on signaling is contained on the dual nodes.
This enables the GNN to capture directional signaling cascades that would be difficult to handle through the convolutional structure of a GNN based on the cell graph only.
\textbf{B.} The main challenges encountered in learning the dynamics of cells in tissues are discreteness, stochasticity, and non-Markovianity (i.e. history dependence).
 These challenges originate from the processes at play in the multicellular dynamics of a biological tissue (panel A).
\textbf{C.} These are addressed by blending three machine learning techniques: graph neural networks (neural networks working on data stored on the vertices and edges of graphs), normalizing flows (generative models that represent complex probability distributions by a sequence of change of variables), and WaveNet (autoregressive models using causal dilated convolutions).
}
\end{figure*}

Unlike the interactions between atoms or molecules, the precise biochemical mechanisms regulating cell dynamics remain largely unknown despite their crucial role in biological processes ranging from morphogenesis to inflammation \cite{Phillips2013}. 
This complexity ultimately arises from the intricate interplay of physical and chemical reactions among myriads of biological molecules inside a cell. 
Schematically, cell dynamics involve three key processes: motion, state transitions, and signaling (Fig.~\ref{fig_gwn}A).
These processes challenge current algorithms, as they require to incorporate stochastic signals with unknown statistics and inherently discrete processes over multiple time scales. 
Moreover, cell interactions are both nonreciprocal and path dependent, since they can be mediated by complex cell signaling cascades. 
These difficulties notwithstanding, rapid advances in imaging routinely allow the automated collection of tissue dynamics datasets with single-cell resolution. 
This raises the prospect of developing machine learning methods for the study of multi-cellular dynamics that infer discrete probabilistic models of tissues~\cite{Rafelski2024,Schoenholz2016b,Berthier2023,Bruckner2024b,Howe2025,Cubuk2015,Sharp2018,Boattini2020,Schmitt2024,Brandstatter2025,Joshi2022,Bruckner2021,Gkeka2020,Schoenholz2016,Frishman2021,Bapst2020,Yang2024,Yamamoto2022}. 
In order to do so, we face  three challenges intrinsic to biological tissues illustrated in Fig.~\ref{fig_gwn}B-C: discreteness,  stochasticity, and non-Markovianity.

First, a natural candidate to describe tissues are graph neural networks (GNN), a variant of neural networks ~\cite{Kipf2017,Scarselli2009,Zhou2020} capable of handling \textit{discrete} data on irregular graphs of varying connectivity through message-passing algorithms that transmit information along their edges.
In a nutshell, the architecture of these neural networks reflects the underlying biology of tissues: cells are represented by nodes in a dynamic graph, and cell signaling is mimicked by the message-passing between the nodes (Fig.~\ref{fig_gwn}A). 
Crucially, the GNN architecture encodes the fact that the units (cells) are indistinguishable from each other through permutation equivariance, and additionally incorporates an implicit bias towards local correlations.
The main practical advantage of this  architecture is to substantially reduce the amount of data required to train compared to convolutional or fully connected deep neural networks that would have to rediscover locality from data.
This requirement can in principle be solved by having more samples, but this is impractical for tissues due to the intrinsic variability of these biological systems.
As the cell state is encoded in the nodes, the isotropic graph convolution underlying standard GNNs is not suitable to capture directed cascades of signaling events that involve correlations between subsequent signaling steps.
The key methodological innovation that allows us to handle this crucial biological complexity is to augment the cell graph with a dual signaling graph where interactions are encoded in nodes (see Fig.~\ref{fig_gwn}A and Methods).
It is this biomimetic architecture of our dual-graph NNs that ultimately captures signaling cascades.

Second, the stochastic dynamics observed in tissues requires modeling probability distributions and transition probabilities. In order to do so, we combine GNNs with normalizing flows, a class of generative models that can be used to efficiently generate, manipulate, and sample complex probability distributions~\cite{Tabak2010,Rezende2015,Kobyzev2021,Liu2019}. 
One of the key deliverables of this approach is the inference of individual cell-level probability distributions from a single tissue-level experiment. 
The reason why this is useful is the following: while it is known that noise can have important implications in biological systems \cite{Amir2018,Golding2024,Hofling2013,Paulsson2005,Tsimring2014}, it is not easy to perform reliable measurements of individual cell level stochasticity within tissues.

Third, the dynamics of biological tissues is often non-Markovian, an inescapable consequence of the fact that we cannot explicitly account for all the processes at play within each cell. 
Hence, one needs to keep track of the tissue over multiple time scales in order to predict its future. 
This is particularly challenging when these time scales are very different, because it requires handling large amounts of data all at once.
In order to tackle this challenge, we take inspiration from WaveNet, a generative model  developed 
to generate natural-sounding speech that mimics human voices~\cite{Oord2016,Wu2019}.
In the case of voice, for instance, the duration of a spoken word is of the order of seconds, while the pitch of sounds can go from 100 Hz to 1000 Hz.
Crucially, however, it is not necessary to resolve the whole audio signal at the shortest time scales. 
It suffices to describe the slowly-varying modulation of a carrier signal, like in AM radio.
In a nutshell, WaveNet combines a version of this multiple time-scale approach, known as a causal dilated convolution, with autoregressive generative models. 
This strategy applied to GNNs is henceforth referred to as graph WaveNets. 

We now go back to biological tissues and provide a step by step guide on how to combine dual-graph WaveNets and normalizing flows into a coherent modeling framework, using the three challenges as organizing principles of our exposition.

\medskip
\noindent\textbf{Tissue dynamics as a neural stochastic differential equation.}
Very much like a pollen grain in water, biological tissues do not evolve in a deterministic fashion.  
The pollen grain can be described by adding a fluctuating noise to the deterministic equation obtained from Newton's laws.
The resulting equation is called a stochastic differential equation.
Here, we model the time evolution of biological tissues by a \textit{neural} stochastic differential equation whose form we sketch using the following pictorial representation:
\onecolumngrid
\begin{equation}
    \raisebox{-0.5\height}{\includegraphics[height=80pt]{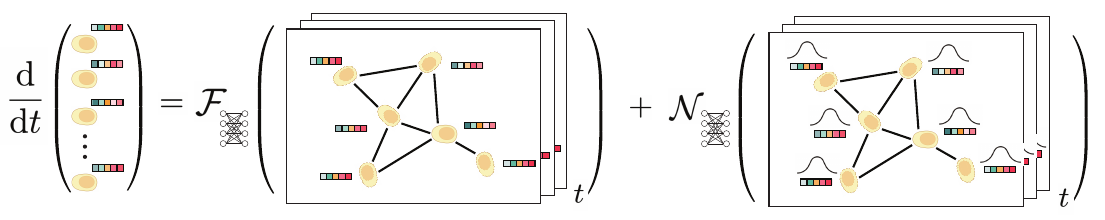}} \label{eq:dyn_eq}
\end{equation}
\twocolumngrid
\noindent It describes the state (\protect\state{}) of each cell (\protect\cell{}) by a set of continuous random variables that can include their position, velocity, and the concentration of certain proteins and genes expressed by the cells globally. 
What makes Eq.~\eqref{eq:dyn_eq} a \textit{neural} equation is that both terms on the right hand side are neural networks. 
The neural stochastic differential equation \eqref{eq:dyn_eq} is composed of a deterministic drift \protect\drift, which may arise from a developmental cell fate, and a fluctuating noise \protect\noise{} which needs not be Gaussian because it could be generated by the internal active dynamics of the cell. 
For instance, in the dynamics of fate decision during cell differentiation, the drift term would represent the gene regulatory network capturing the complex dynamics among numerous transcription factors~\cite{Karlebach2008}. 
The noise term would capture the intrinsic stochasticity of chemical reactions in the process of gene expression and DNA binding events of transcription factors.

In order to describe individual cells proliferating and dying, we encode these discrete data in a graph whose varying connectivity is determined by the time-dependent interaction network. 
The discreteness challenge (first row in Fig. \ref{fig_gwn}B-C) is tackled by choosing the neural network \protect\drift{} in Eq.~\eqref{eq:dyn_eq} to be a graph neural network whose connectivity evolves in time due to cell motion, division, and removal, so the structure of the locality bias is dynamic and informed directly by the data.

\begin{figure*}
\vspace*{-1.25cm}
\includegraphics[width=0.9\textwidth]{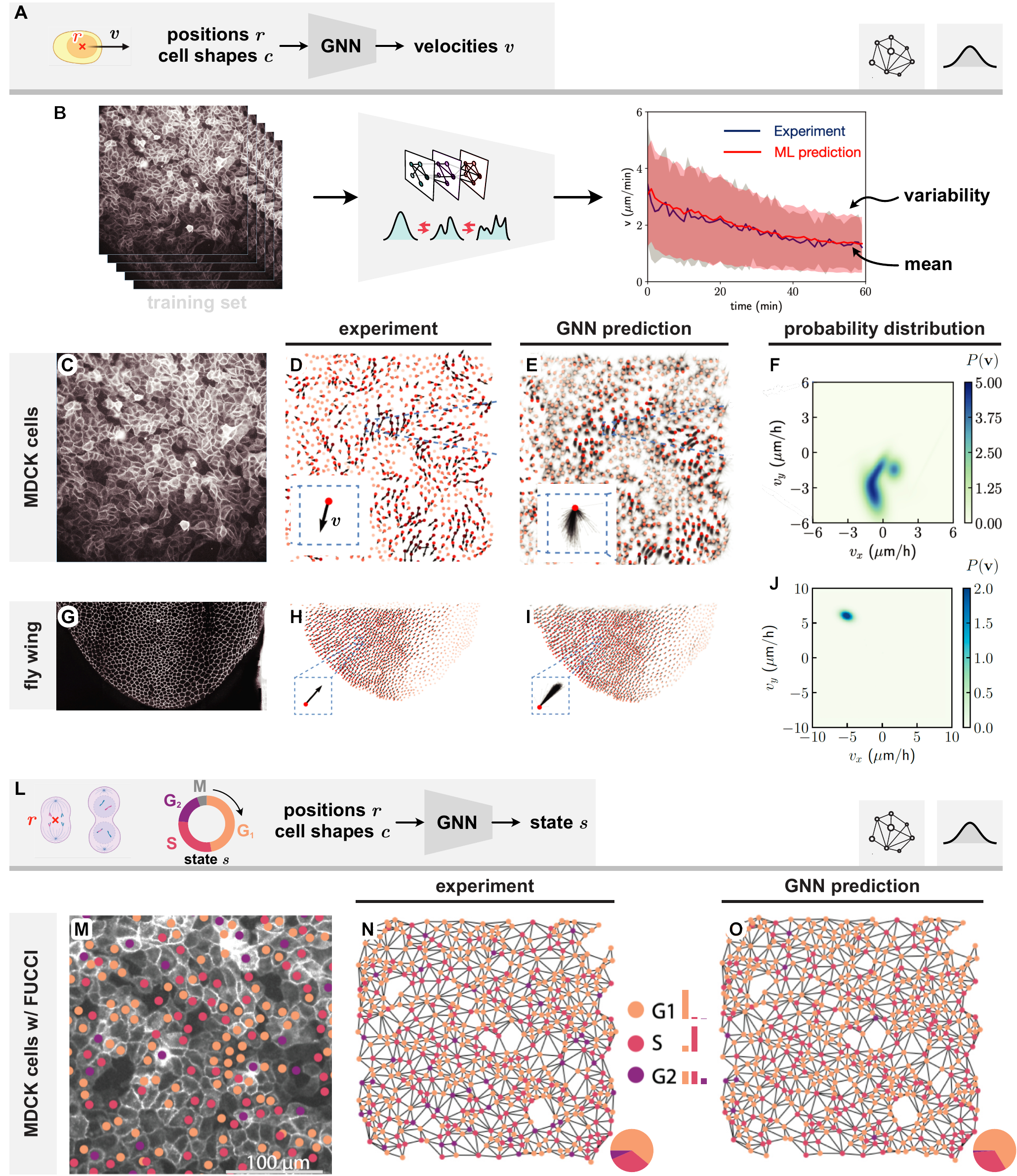}
\caption{\label{fig_migration_division} \textbf{Learning cell dynamics.} \textbf{A.} We aim at predicting the velocity $\bm{v}_i$ of all individual cells from their positions $\bm{r}_i$ and cell shapes and areas ($c_i$) in tissues.
This requires handling discrete and stochastic data, but no non-Markovian effect.
\textbf{B.} The ML pipeline allows us to predict the evolution of both average features (red and blue lines)  and noise (manifested through observed variability; red and blue regions).
\textbf{C-F.} We apply the technique to a monolayer of epithelial MDCK cells, which undergo stochastic migration and division in a confined environment (panel C). 
Cells become more rounded as they divide and gradually transit from a fluid state into a glassy state.
Their motility, measured by the spatially averaged cell speed, decreases over time. 
Our ML model not only captures the same decreasing trend (panel B), but also accurately predicts the spatial variation (shaded region). 
Panel D shows the velocity field at single-cell resolution. Even though an experimental movie only provides a particular realization of the stochastic dynamics, the probabilistic design enables us to infer the underlying statistics through maximum likelihood optimization (panel E). This allows us to further extract the ensemble properties of individual cells (panel F). 
\textbf{G-J.} We applied our ML algorithm to study the growth of a fly wing~\cite{Etournay2015,Etournay2016}. 
It automatically recognizes the deterministic nature of such a developmental system. The predicted probability distribution of cell migration velocity is nearly a $\delta$-function (panel J). 
Note that all the ML predictions shown in this paper are produced on the test dataset, which has never been seen by the ML model during training.
\textbf{L.} We aim at predicting the state of individual cells within the cell division cycle from the individual positions in the tissue. 
In the division process, the cells go through three states, $G_1$, $S$, and $G_2$ and then undergo mitosis ($M$).
It is a stochastic process discrete in time meanwhile subject to regulations from neighboring cells through mechanical stresses, but here we do not consider non-Markovian effects.
\textbf{M.} In a monolayer of MDCK cells, we can identify the division states of individual cells at any given time. \textbf{N-O.} Comparison between experimental ground truth (panel N) and ML predictions (panel O).  After training, our model can predict the cell division state with accuracy over 80\%.
}
\end{figure*}

Stochastic effects, our second challenge (second row in Fig. \ref{fig_gwn}B-C), are captured by the noise term \protect\noise{} in Eq.~\eqref{eq:dyn_eq}.
In particular, we must allow the noise in the states of individual cells to be correlated with each other through the interaction network. 
In order to model the joint probability distribution of all the cells in the tissue, we resort to our dual-graph normalizing flow architecture. We enforce that different cells should behave identically, with the same drift term and the same probability distribution of the noise, when they have the same biological state \protect\state{} and the same local environment (i.e. other cells in the graph \protect\graph{}), like real cells tend to do. 
The key deliverable of our biomimetic algorithms is that they are capable of inferring single-cell-level probability distributions (that can be then sampled to compute averages and correlations), despite using only a reduced amount of data that would otherwise not suffice to reliably estimate these quantities. 

To capture memory effects, arising from our third challenge non-Markovianity (third row in Fig.~\ref{fig_gwn}B-C),  we allow \protect\drift{} to be a function of the system history (represented by the time sequence \protect\seq) which describes how the states and interactions networks of each cell (represented by the graph \protect\graph) vary over time. 
Similarly, the noise is sampled from a joint probability distribution (represented by the symbol \protect\prob), which is also history dependent and allows correlations between cells generated through their interaction network. 
We encode long-term memory effects through causal dilated convolution layers, which perform multi-scale context aggregation, following the approach used in WaveNet \cite{Oord2016,Yu2015,Dutilleux1990,
Holschneider1989}. 

\begin{figure*}
\includegraphics[width=0.9\textwidth]{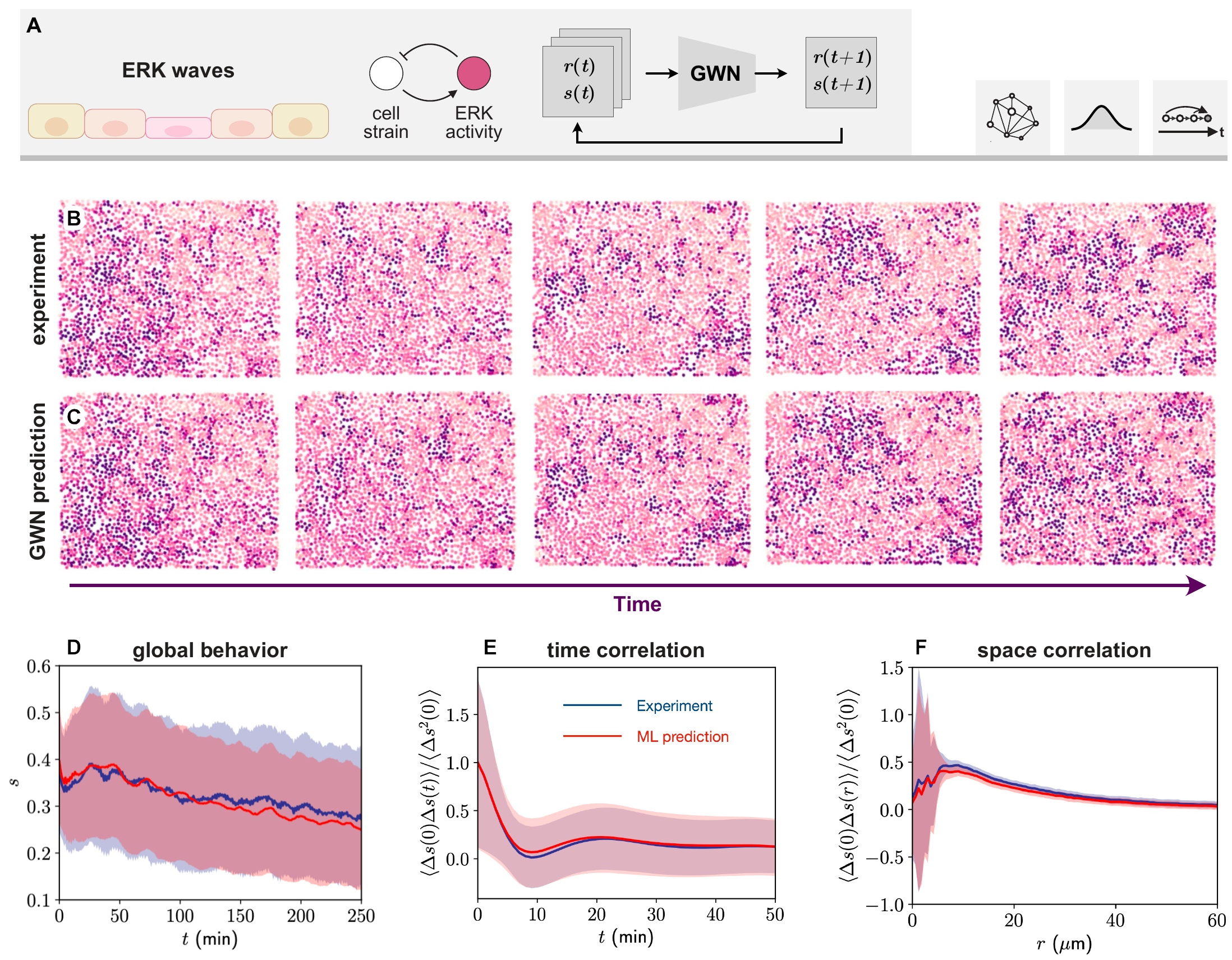}
\caption{\label{fig_signaling} 
\textbf{Learning dynamic cell signaling.} \textbf{A.} By sending out ERK signals, the cells can coordinate their movements (such as lateral expansion and contraction), leading to collective migration of the cells through a feedback loop between ERK activity and cell strain.
Here we perform a ML study of the propagation of ERK waves where we aim at predicting the future of the system from its past (the state of the system is represented by the positions $\bm{r}(t)$ of the cells as well as their internal states $\bm{s}(t)$). 
In addition to the challenges of discreteness and stochasticity already present in the other examples, here we need to take non-Markovian effects into account.
Given this long-term correlation in ERK waves, here we integrate our graph neural network with an advanced sequence model called WaveNet. These two networks work seamlessly in our architecture: the WaveNet encodes the environmental changes of individual cells over time, whereas the graph neural networks collect and redistribute such sequential information among each cell and its neighbors. 
\textbf{B-C} Comparison between experimental ground truth (panel B) and ML prediction (panel C) of ERK signals. 
\textbf{D-F.} Our ML model accurately predicts the overall decreasing trend of the spatially averaged ERK intensity $s$ (panel D) as well as its spatial variations (shaded region).
Furthermore, our model also captures the right statistics of the ERK waves: the predicted time and spatial correlation functions (panels E and F) match well with the experimental ground truth.
In panels D-F, the ML prediction is shown in red while the experimental ground truth is shown in blue.
}
\end{figure*}

We now show that our algorithmic framework allows us to efficiently learn both terms \protect\drift{} and \protect\noise{} of Eq.~\eqref{eq:dyn_eq} at the same time from experimental movies. We present our examples in order of increasing computational difficulty starting from situations where the random term \protect\noise{} dominates (epithelial tissues, Fig.~\ref{fig_migration_division}A-F) or where the deterministic drift \protect\noise{} dominates (development, Fig.~\ref{fig_migration_division}G-J) culminating in the most challenging case of mixed stochastic and deterministic dynamics (cell signaling mediated by ERK waves, Fig.~\ref{fig_signaling}).

\medskip
\noindent\textbf{Noisy cell dynamics in epithelia.} As a first application of our algorithmic framework, we proceed to tackle a basic biophysical question that is still open: is the \textit{noisy} motion of cells determined (i.e. predictable) solely by their geometry and position within the tissue? In this case, the deterministic drift term \protect\drift{}  in Eq. \eqref{eq:dyn_eq} is negligible and the dynamics is dominated by the fluctuating noise \protect\noise{} that is approximately Markovian, making the addition of the WaveNet part unnecessary in this simpler example.
Prior models and experiments have suggested that a correlation exist between cell geometry and tissue flow but only for average cell motions \cite{Bi2015,Bi2016,Park2015,Mongera2018,Petridou2021,Devany2021}.

Instead we task our GNNs with predicting the stochastic dynamics of {\it each} cell in the tissue at every instant (Fig.~\ref{fig_migration_division}A). 
What makes this noisy cell motion hard to predict is that it is  generated by intracellular active forces ultimately traceable to stochastic processes within the cytoskeleton not easily  accessible experimentally. This biohysical hurdles notwithstanding, we can predict a probability distribution for the displacement of each cell based on the current configuration.
Training is done for each system of interest which allows for the neural network to determine the level of stochasticity exhibited by cells in the tissue.

We apply our algorithms to epithelial tissue mono-layers of Madin-Darby canine kidney (MDCK) cells, a well studied model system for collective cell migration \cite{Aoki2017,Rorth2009,Hakim2017} for which large amounts of data are available for training (Fig.~\ref{fig_migration_division}B-F). 
We performed time lapse imaging experiments taking images of cell membranes and nuclei every 10 minutes over roughly 24 hours (Fig.~\ref{fig_migration_division}C-D), see Methods for details. The cell membrane and nuclei were then segmented and tracked to produce a list of cell positions, sizes, shapes and displacements to input into the GNN. 
As it is typically not possible to correctly segment and track every cell, we have developed an automated method for filling in missing cells with predicted "dummy cells" that have the correct average properties.
Figure~\ref{fig_migration_division}D shows an example of the resulting data: cell centers are shown as points colored by speed and their velocities are represented by arrows. 
We train the network with all these inputs and task it to make a prediction of the displacement from the other geometric variables.

We now attempt to discover the \textit{full} probability distribution of cell speeds using our generative model even if a direct measurement of this quantity is not possible in our small dataset without the implicit data augmentation enabled by our equivariant ML architecture (Fig.~\ref{fig_migration_division}A-C). Inspection of Fig.~\ref{fig_migration_division}B reveals that at each time point,  the prediction (red) made by our neural networks fully matches the experimental data (blue), not only for the mean (continuous line) but also the standard deviation (shaded area) of cell speed.
We emphasize that this prediction is made only from cells geometry and positions.
Observing the displacements of each individual cell, we notice that the GNN is able to predict the small scale coherent flows seen in cell monolayers (Fig.~\ref{fig_migration_division}D-E) \cite{Bi2016}. 
Furthermore, we see the stochastic behavior of cells by the various cell displacements predicted by the GNN. 
Here, the predicted probability distribution (Fig.~\ref{fig_migration_division}F) has multiple dominant peaks suggesting that for the current configuration, either of these displacements is favorable depending on forces applied by individual cells within tissues.

\medskip
\noindent\textbf{Deterministic cell dynamics in morphogenesis.}
We now move on to a different example where we compare and contrast the results presented above with predictions made on a developing fly wing (Fig.~\ref{fig_migration_division}G-J).
To do so, we applied our GNN analysis to published fly wing data \cite{Etournay2015}.
In this case, the deterministic drift term \protect\drift{} in Eq.~\eqref{eq:dyn_eq} dominates and it is the noise \protect\noise{} that is small. While predictive models at the cell scale for such systems are still lacking, developmental systems show more stereotypical behavior across embryos amenable to continuum modeling approaches \cite{Streichan2018,Lefebvre2023,Hannezo2019,Garcia2015,Saadaoui2020,Streichan2018,Tavano2025,Bruckner2024}.
Our algorithm successfully predicts motion across the wing despite only being trained on a subset of a single experiment~\cite{Etournay2015,Etournay2016} (compare Fig.~\ref{fig_migration_division}H and I). 
We observe that the predicted distribution of displacements are noticeably more deterministic than the monolayers in vitro showing a single strong peak in the probability distribution (Fig.~\ref{fig_migration_division}J, compare with the in vitro monolayer in panel E). 
This matches our intuition that the developmental system should behave in a reproducible manner while the MDCK monolayer may be much more chaotic.
We emphasize that the inferred single-cell probability distributions in Fig.~\ref{fig_migration_division}F and J correspond to a single experiment (plus an already-trained network). 
This ability to infer ensemble quantities from a single realization ilustrates a key practical strength of our approach.

\medskip
\noindent\textbf{Cell state transitions in the cell cycle.} Next, we attempt to predict the internal biological state of the cell from cell geometry only. 
We focus on the cell cycle, a sequence of proliferation and division composed of four phases (Fig.~\ref{fig_migration_division}L): the $G_1$ phase, a growth phase prior to DNA replication; the $S$ phase, when DNA is replicated, and the $G_2$ phase, a second growth phase when the cell prepares to enter mitosis ($M$). 
The $G_1$ phase has the largest variation in duration, which is determined by whether conditions are met to proceed to the $S$ phase.  
By contrast, in MDCK cells, the $S$ and $G_2$ phases are typically 8--10 and 2--4 hours, respectively.
In epithelial monolayers geometric variables are known to correlate with the cell cycle. 
For example, the number of neighbors a cell has depends on the relative time since dividing compared to other cells \cite{Gibson2006,Gibson2009}, and the area of a cell can be used to infer if it will become cell cycle arrested due to contact inhibition \cite{Devany2023}. However, a comprehensive model which can make accurate predictions at the single cell level does not exist. We asked the GNN to integrate these geometric information to make a prediction of the cell state. 
This is challenging because of the stochastic nature of the cell state transitions between phases of the cell cycle, the discreteness of signaling states and the relationship between the properties of neighboring cells.

We analyzed a previously published dataset of time lapse images of MDCK cells expressing a membrane marker and fluorescent cell cycle reporter (FUCCI) \cite{Devany2023}, see Fig.~\ref{fig_migration_division}M. 
From these, we determined the position, area, shape, displacement and cell cycle state of each cell (Fig.~\ref{fig_migration_division}N). 
We use this information to train the model to make predictions of the cell cycle state from geometric information only.
Notably, this is enough for the network to make an accurate prediction of the cell cycle in over 80$\%$ of cells in the monolayer (Fig.~\ref{fig_migration_division}O, compare with panel N; the bar charts on the right of the legend give the conditional distribution of the GNN prediction conditioned on the ground truth state being $G_1$, $S$, or $G_2$). 
We note that the GNN underpredicts the state $G_2$.
This observation suggests that there is not enough biophysical information in the cell geometry to determine whether a cell is in the $G_2$ phase. Note that we cannot rule out sample under representation in our data set (see pie charts in inset).

\medskip
\noindent
\textbf{Dynamic cell signaling in ERK waves.}
Finally we deploy our algorithms in situations where the propagation of cell signaling across a tissue is stochastic, a bit like turbulent waves at sea. 
In these cases, both the deterministic drift \protect\drift{} and the noise \protect\noise{} in Eq. \ref{eq:dyn_eq} are equally important.
The activation of a cell causes a signal to be passed to neighbor cells which
activates them and in turn propagates the signal forward. 
Predicting such processes requires knowledge of the history of the system (represented by a stack of frames in Fig.~\ref{fig_signaling}A). 
In this case, the combination of GNN with WaveNet (GWN) is necessary because conventional GNNs fail. 
At each layer of the neural network, a GNN is used to encode the spatial dependencies between nodes in the graph, while the dilated convolutional layers capture the temporal dependencies between samples. 
The combination of these two types of neural network layers allows the model to predict the propagation of ERK (extracellular signal-regulated kinase) signaling, a signaling pathway involved in the regulation of cell division in differentiated cells~\cite{Aoki2013,Hino2020,Ram2023}, see Fig.~\ref{fig_signaling}A.
Recently, ERK activity sensors have been developed and it was discovered that ERK signaling produces mechanochemical waves that propagate through a tissue~\cite{Boocock2020,Hino2020,Boocock2023}. 
However, models of this process are limited by the challenge of simultaneously predicting the stochastic ERK signals and the motion of cells, which occur at the same timescale. We are not aware of models that can quantitatively reproduce both the average evolution and the stochastic correlations, nor inferring these quantities from data in a model agnostic manner. Here we ask: 
can we bridge this gap with our dual graph WaveNet framework.

We use a previously published dataset of ERK dynamics in epithelial MDCK monolayers and segment the cells to track the position, nuclear morphology and ERK signaling state of each cell~\cite{Boocock2020}. 
We trained the GWN on a time series of this data, and then provided it with an unseen test set. 
The first 7 frames of the test set are provided to the GWN because history is required to make further predictions. 
Next, we query the model to predict the future signaling dynamics on unseen samples. 

We find that the GWN produces an output of wavelike dynamics that mimic the experiment (Fig.~\ref{fig_signaling}B-C). 
As the propagation of these waves is highly sensitive to the initial conditions and noise in the system, the exact wave pattern cannot be predicted out to long time scales. 
Nonetheless, the predicted signaling patterns match the experiment in magnitude and fluctuations (Fig.~\ref{fig_signaling}D). 
Further, when we analyze the spatial and temporal correlations of the signaling waves we see that both the magnitude and variation in these correlations match the experiment (Fig.~\ref{fig_signaling}E-F). 
This suggests that the same variables (cell morphology and positions plus ERK activity) that are relevant to the deterministic dynamics modeled in Ref.~\cite{Boocock2020} (i.e., that allow one to predict the future of the system) control the stochastic dynamics of the system.

Beyond this qualitative conclusion, our algorithm reveals the different time scales at play in the tissue: 
Fig.~\ref{fig_signaling}D shows evolution on a long time scale of the order of hundreds of minutes
while Fig.~\ref{fig_signaling}E shows how the network captures the two shorter time scales associated with the oscillation and decorrelation of the waves, of the order of 10 to 20 minutes. 
It is the interplay between the protein signaling dynamics on short time scales and the accumulation of many discrete events (cell division and rearrangement) that make the tissue move on the longer time scales.
To sum up, our generative model discovers directly from data the mechanochemical coupling between active stress and the ERK pathway that underlie the spatiotemporal patterns in the MDCK monolayers and gives access to all many-body correlations.

\medskip
\noindent
\textbf{Outlook.}
One of the current major challenges in biology is to measure gene expressions and regulatory landscapes of individual cells, and to integrate this information with spatial data describing the environment of the cell within tissues and organs~\cite{Bressan2023,Levsky2002,Satija2015,Wolf2018,Kulkarni2019,Haviv2024}.
Our work paves the way towards a holistic description of tissues, through a joint probability distribution, rather than starting from the gene expressions of isolated cells. By integrating this multicellular context with single-cell gene expression data, our approach could directly yield the reduced representation of cell state transitions in a way that captures multiscale correlations from the get-go.

Furthermore, our trained neural networks could be used as a digital twin in clinical studies of tissue inflammation. This would entail first calibration with lab data, where many biological markers are measured, and subsequently deployment on patient data, even if only a more limited amount of measurements is available.

\medskip
\noindent\textbf{Acknowledgements.}
We thanks Daniel Seara, Doruk Efe Gökmen, and Smayan Khanna for critical feedback on the manuscript.
This research was partly supported from the National Science Foundation through the Physics Frontier Center for Living Systems (PHY-2317138) as well as NSF (DMS-2235451) and Simons Foundation (MPS-NITMB-00005320) to the NSF-Simons National Institute for Theory and Mathematics in Biology (NITMB). M.G. and V. V. are Chan Zuckerberg Biohub Chicago Investigators. This work was completed in part with resources provided by the University of Chicago’s Research Computing Center.
M.F. and V.V acknowledge partial support from the France Chicago center through a FACCTS grant. 

\clearpage

\section*{Methods}

\subsection{Machine learning design for biological interactions}

From swarms of E. coli to clusters of epithelial cells, living systems exhibit a vast variety of collective motions. This largely stems from the dynamic interaction between individual cells, which can be time-varying, non-conservative, and even non-reciprocal. Despite such richness, these interactions often share two common natures: locality and universality. Consider the example of multicellular tissues, where cells engage primarily with their nearest neighbors. This local interaction is mediated through mechanical forces and chemical signals exchanged across shared membranes.  In addition, cells of the same type generally adhere to a consistent principle of how to sense and respond to environmental stimuli.

We use graph convolution networks (GCNs) to leverage the locality and universality of interactions to focus on learning the complex interaction patterns among cells. Within this framework, a graph acts as a generic representation of a living many-body system, with individual cells forming a typically irregular network. GCN applies the same neural network to each cell, meanwhile, uses graph convolution to receive and send information with its neighbors. By using shared parameters, it is designed not only to extract the general principles in local interactions between cells but also to greatly reduce model complexity without sacrificing expression power, making it possible to train deep learning models with limited experimental data.

In the overarching design, our model utilizes an encoder-decoder architecture, see Fig.~\ref{fig:dual_graph2}b. The encoder mimics the sensing process. It uses a combination of GCN and WaveNet to gather information on the local environment of cells by embedding the spatio-temporal behaviors of individual cells and their neighbors. The decoder models the response process. It is represented as a generalized Langevin equation Eq.~\eqref{eq:dyn_eq}, which simultaneously derives both the deterministic drift and stochastic noise based on the spatiotemporal embeddings of the cells from the encoder. Considering that the stochastic noise can be non-white and correlated across different cells, we integrate GCNs with normalizing flow to capture such richness. This allows us to directly infer 
the joint probability distribution of all cells conditional on their local environment. Detailed descriptions of each machine learning module are elaborated further below.

\vspace{5mm}

\subsection{Dual-graph convolution networks}

The off-lattice arrangement of a living many-body system is naturally represented by a cell graph $\Gc = \big(\Vc, \Ec\big)$. It is composed of a set of nodes $\Vc = \big\{n_i^\comp \,|\, n_i^\comp \in \Gc \big\}$ denoting individual cells indexed by $i$ and a set of edges $\Ec = \big\{e_{ij}^\comp \,|\, e_{ij}^\comp \in \Gc \big\}$ denoting the interconnectivity defined as all the cell pairs $(n_i^\comp, n_j^\comp)$ that undergo direct interactions. 

The conventional implementation of GCN can be summarized as a message-passing process often implemented in an auto-regressive manner. Taking the input vectors as initial states, $\h^{(0)}(n_i) = \x(n_i)$ and $\h^{(0)}(e_{ij}) = \x(e_{ij})$, GCN progressively encodes nodes and edges by constructing messages $\h^{(l+1)}(e_{ij})$ along incoming edges then gathering them for the encoding of nodes $\h^{(l+1)}(n_{i})$
\begin{align}
    \h^{(l+1)}(e_{ij}) &= M\Big[\h^{(l)}(n_i), \, \h^{(l)}(n_j), \, \h^{(l)}(e_{ij})\Big], \label{eq:gcn-1} \\
    \h^{(l+1)}(n_i) &= U\Big[\h^{(l)}(n_i), \sum_{j\in \N(i)} \h^{(l+1)}(e_{ij}) \Big], \label{eq:gcn-2}
\end{align}
where $M$ and $U$ are neural networks such as multi-layer perceptron (MLP), index $l$ denotes $l$-th round of graph convolution, and $\N(i)$ is the neighbor set of node $i$.

\begin{figure*}
\includegraphics[width=0.9\textwidth]{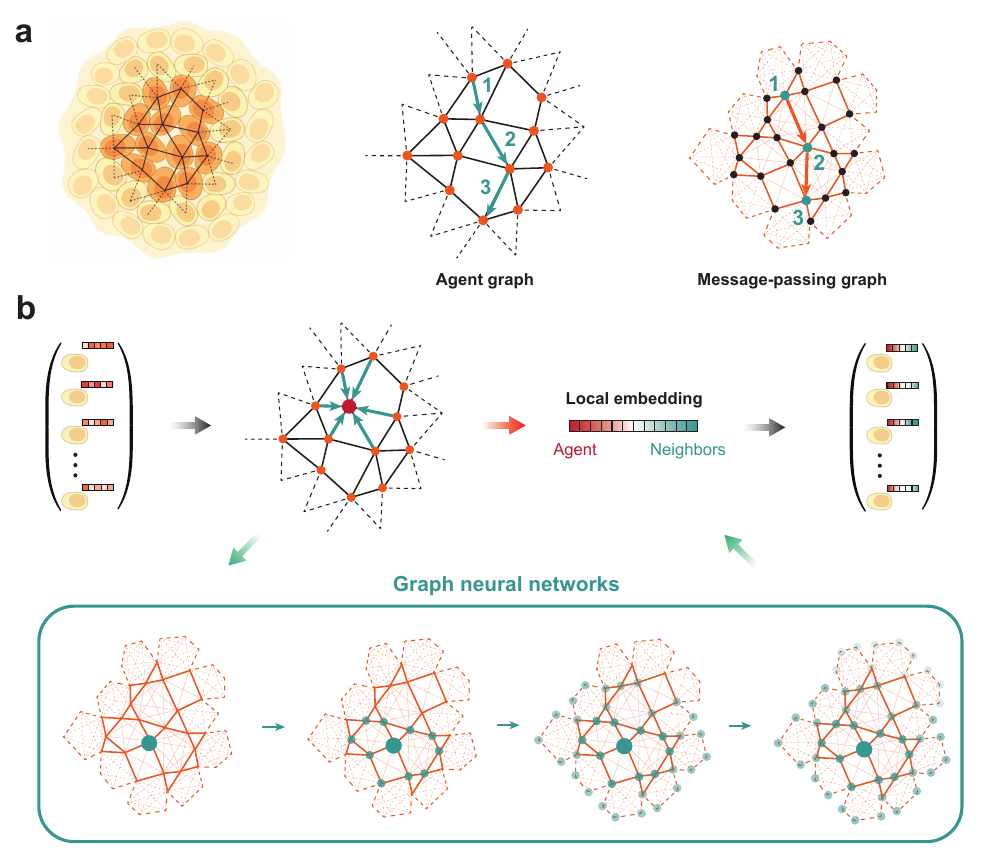}
\caption{\label{fig:dual_graph2} \textbf{Graph neural networks.} \textbf{a.} Dual graph representation. We represent the spatial arrangement of the agents as an agent graph $\mathcal{G}^\text{a}$, where nodes $n^\text{c}_i$ correspond to cell nuclei and edges $e^\text{c}_{ij}$ indicate the connectivity between two adjacent cells sharing a common membrane. To allow complex cell-cell interactions like signaling, we further employ a message-passing graph $\mathcal{G}^\text{m}$, which form a dual structure in relation to the cell graph: $n^\text{m}_\mu \leftrightarrow e^\text{c}_{ij}$ and $e^\text{m}_{\mu\nu} \leftrightarrow \left(e^\text{c}_{ij}, e^\text{c}_{jk}\right)$. A signal cascade causes statistical dependency of the $\mathcal{G}_\text{c}$ edges in a sequential manner, e.g. $e_{ij} \to e_{jk} \to e_{kl}$, which is converted into the dependency of the $\mathcal{G}_\text{m}$ nodes $n_{\mu} \to n_{\nu} \to n_{\xi}$. The latter can then be explicitly encoded by graph convolution. \textbf{b.} Graph convolution. Information of individual cells and any explicit relations with its neighbors are passed onto the nodes and edges of the cell graph $\mathcal{G}^\text{c}$, respectively. Then we duplicate the same information on the message-passing graph $\mathcal{G}^\text{m}$ and encode how each cell-cell relation depends on local environment using a sequence of graph convolutions. The obtained embeddings are returned back to the corresponding $\mathcal{G}_\text{c}$ edges, capturing the relation between any given cell and its neighbors in a directional fashion. Together with the information of the cell, they provide a complete spatial embedding on the local environment, which can then be used in other ML modules.}
\end{figure*}

This design explicitly encodes the dependency between nodes but misses the dependency between edges. However, cells could undergo complex interactions that involve the latter. For instance, cell signaling in a tissue is a cascading phenomenon that often exhibits statistical dependence between sequent signals, see Fig.~\ref{fig:dual_graph2}a.

To address this technical challenge, we further construct a message graph $\Gm = \big(\Vm, \Em\big)$, which contains a node set $\Vm = \big\{n_\mu^\mess \,|\, n_\mu^\mess \in \Gm \big\}$ and an edge set $\Em = \big\{e_{\mu\nu}^\mess \,|\, e_{\mu\nu}^\mess \in \Gm \big\}$. $\Gm$ and $\Gc$ display a dual correspondence: 
\begin{align}
    n_\mu^\mess &\leftrightarrow e_{ij}^\comp, \\
    e_{\mu\nu}^\mess &\leftrightarrow \big(n_{i}^\comp, n_{j}^\comp, n_{k}^\comp\big).
\end{align}
Here edges $e_{ij}^\comp$ in the cell graph $\Gc$, which mark all the possibilities of single-step message passing, are mapped to corresponding nodes $n_\mu^\mess$ in the message graph $\Gm$. 
Edges $e_{\mu\nu}^\mess$ in $\Gm$ correspond to edge tuples $(e_{ij}^\comp, e_{jk}^\comp)$ in $\Gc$, which is equivalent to sequential node paths $n_i^\comp \to n_j^\comp \to n_k^\comp$ that mark all the possibilities of two-step message passing.
This dual-graph design transforms edge dependence in $\Gc$ to node dependence in $\Gm$. Note that unlike $\Gc$, $\Gm$ is not a bi-directional graph: a reversed edge $e_{\nu\mu}^\mess$ in $\Gm$ leads to a invalid edge tuple  $(e_{jk}^\comp, e_{ij}^\comp)$, which does not correspond to any sequential node path in $\Gc$.

Our graph convolution is developed on the dual-graph representation, see Fig~\ref{fig:dual_graph2}b. Input states of individual cells and their interconnectivity are passed into the nodes and edges of the cell graph $\Gc$, respectively. They are then used to prepare the input states for $\Gc$:
\begin{align}
    \x(n_{\mu}^\mess) &= F\Big[\x(n_i^\comp), \x(n_j^\comp), \x(e_{ij}^\comp)\Big], \\ \x(n_{\nu}^\mess) &= F\Big[\x(n_j^\comp), \x(n_k^\comp), \x(e_{jk}^\comp)\Big], \\
    \x(e_{\mu\nu}^\mess) &= G\Big[\x(n_i^\comp), \, \x(n_j^\comp), \, \x(n_k^\comp)) \Big],
\end{align}
where $F$ and $G$ are trainable functions modeled by MLPs. 
A sequence of GCNs, operations defined as Eqs.~(\ref{eq:gcn-1}-\ref{eq:gcn-2}), are applied to the message graph $\Gm$, with embedding vectors produced by each GCN fed into the subsequent one as input states.
Each GCN operation expands the receptive field of individual nodes by one layer of neighbors, see Fig~\ref{fig:dual_graph2}b. 
The ultimate node embedding in $\Gm$ is taken as message in $\Gc$:
\begin{equation}
    \m(e_{ij}^\comp) = \h(n_\mu^\mess).
\end{equation}
where $\m(e_{ij}^\comp)$ encodes all the information that are passed to node $n_j^\comp$ through node $n_i^\comp$ but can be generated by further neighbors beyond $n_i^\comp$. Finally, we encode the local environment of each cell as
\begin{equation}
    \h(n_{i}^\comp) = V\Big[\x(n_i^\comp), \sum_{j\in \N(i)} \m(e_{ij}^\comp) \Big],
\end{equation}
which includes the input states of node $\x(n_i^\comp)$ and all the messages from its neighbors, with $V$ as a MLP function.

\vspace{1cm}

\subsection{Graph WaveNet}

The behavior of individual cells depends on the past history of their local environment. It includes the evolution of the cells and their neighbors quantified by state vectors $\x_i^\comp(t)$ as well as the variation of their interconnectivity represented by cell graph $\Gc(t)$ over time. To encode such spatiotemporal behavior, we need a method that can properly assemble $\x_i^\comp(t)$ and $\Gc(t)$ at each timestep in the past.

Here we employ a combination of Graph Neural Networks and WaveNet, similar to Ref.~\cite{Wu2019}. 
As for pre-processing, we first apply the dual-graph convolution to obtain the initial embedding of individual cells at each time step, denoted as $\h(n_i^{\comp},t)$. Then a $T$-steps-long history of any given node $n_i^\comp$ and its local environment can be represented as a stack of such embedding vectors:
\begin{equation}
    \h^{(0)}(n_i^{\comp},t)=
    \bigl[\h(n_i^{\comp},t),\,\dots,\,\h(n_i^{\comp},t-T+1)\bigr]^{\mathsf T}.
\end{equation}

\begin{figure*}
\includegraphics[width=0.6\textwidth]{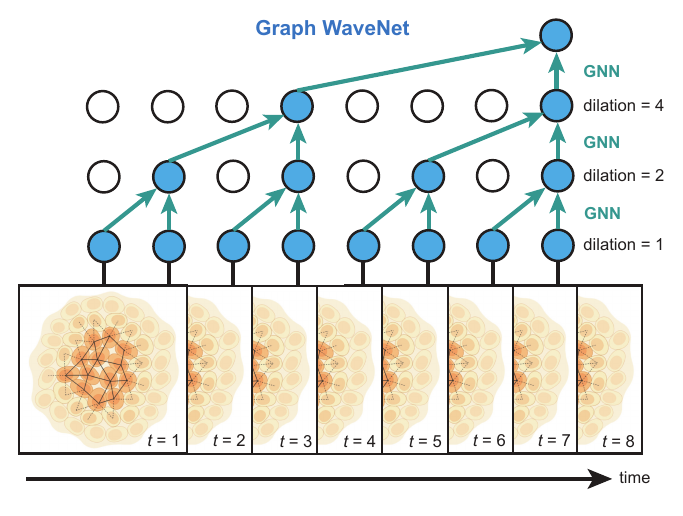}
\caption{\label{fig_gwn_methods} \textbf{Graph WaveNet.} The standard WaveNet uses a stack of 1D dilated convolutions to process sequential data. The receptive field of WaveNet grows exponentially with the depth of the network (see the tree of blue nodes), allowing us to handle long-term memory effects. Here we integrate graph neural networks into each dilated convolution (see green arrows), to further incorporate spatial relation between agents. The resulting model can therefore provide the spatiotemporal embedding of each individual agents, which encode how its local environment varies in the past.}
\end{figure*}

To efficiently capture long-term memory effects, $\h^{(0)}(n_i^{\comp},t)$ is processed by a cascade of $D$ dilated
one-dimensional convolutions along the temporal axis. At level $d$ $(=1,\dots,D)$, we apply a convolution of kernel width $k=2$ and
dilation $2^{d-1}$,
\begin{equation}
    \widehat{\h}^{(d)}(n_i^{\comp},t)=
    \text{CausalConv1D}_{k,\,2^{\,d-1}}
    \!\bigl(\h^{(d-1)}(n_i^{\comp}, t)\bigr),
\end{equation}
where a casual design (Fig.~\ref{fig_gwn_methods}) is employed to guarantee that prediction at any given time step only depends on past and present inputs, not future ones.

At each level, the embeddings of individual nodes are shared laterally with their neighbors via the cell graph $\Gc(t)$:
\begin{align}
    \h^{(d)}(e^\comp_{ij}, t) &= M\Big[\widehat{\h}^{(d)}(n_i^{\comp},t), \, \widehat{\h}^{(d)}(n_j^{\comp},t) \Big], \label{eq:gcn-1m} \\
    \h^{(d)}(n^\comp_i, t) &= U\Big[\widehat{\h}^{(d)}(n_i^{\comp},t), \sum_{j\in \N(i, t)}  \h^{(d)}(e^\comp_{ij}, t) \Big]. \label{eq:gcn-2m}
\end{align}
where $\mathcal N(i,t)$ is updated at every step to respect the evolving network.

Thanks to dilation, Each successive level doubles the temporal reach of the filter while keeping the parameter count fixed. After the final dilation the receptive field spans $T = k\,(2^{D}-1)$, easily covering the decay times of mechanical and signaling correlations observed in experiments.

The embedding vector obtained from Graph WaveNet
\begin{equation}
    \mathbf{z}(n_i^{\comp},t)=\h^{(D)}(n_i^{\comp},t)
\end{equation}
captures both \emph{where} and \emph{when} cell $i$ interacted over the
last $T$ steps. 

\begin{figure*}
\includegraphics[width=0.8\textwidth]{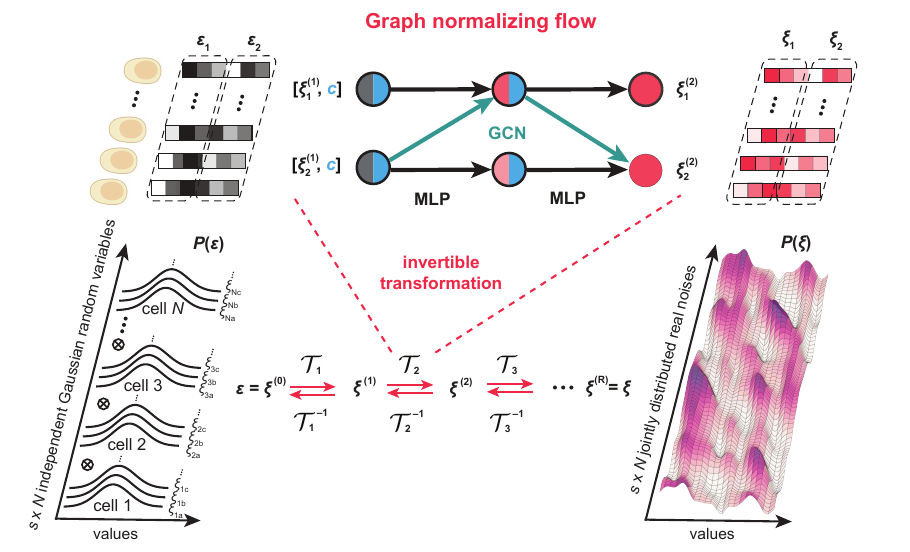}
\caption{\label{fig_gnf} \textbf{Graph normalizing flow.} A multi-dimensional random variable $\mathbf{n}$ can potentially follow a complex joint probability distribution $\mathbf{n} \sim P(n_1, n_2, \cdots, n_m)$. By applying a series of invertible variable transformations, a normalizing flow allows us to establish an one-to-one mapping between $\mathbf{n}$ and white noises $\mathbf{\xi}$, which follows independent normal distributions $\mathbf{\xi} \sim \mathcal{N}(\xi_1)\mathcal{N}(\xi_2) \cdots \mathcal{N}(\xi_m)$. Here we modify the standard normalizing flow by integrating graph neural networks into each transformation step and allowing it to take the spatiotemporal embedding of individual agents as input. The resultant graph normalizing flow can then predict the joint probability distribution over all the agents conditional on the history of their local environment.}
\end{figure*}

\subsection{Graph normalizing flow}

Fluctuations in living matter are highly structured. For example, force bursts can travel through neighboring cells, producing
stochastic forces that are coupled in space.
We denote the instantaneous noise by
\begin{equation}
    \boldsymbol{\xi}(t)=
    \bigl[\boldsymbol\xi(n_1^\comp, t),\dots,\boldsymbol\xi(n_N^\comp, t)\bigr]^\mathsf{T},
\end{equation}
and seek its joint probability density \emph{conditioned} on a set
$\mathbf{c}(t)$ of node-wise embeddings:
\begin{equation}
    \mathbf{c}(t)= 
    \begin{cases}
\bigl[\h(n_1^\comp, t),\dots,\h(n_N^\comp, t)\bigr]^\mathsf{T}
        & \text{state conditioning},\\[6pt]
        \bigl[\z(n_1^\comp, t),\dots,\z(n_N^\comp, t)\bigr]^\mathsf{T}
& \text{history conditioning}
    \end{cases}\label{eq:C_def}
\end{equation}
where the bold $\mathbf{h}$ come directly from the dual-graph encoder
and encode only the present frame, whereas the bold
$\mathbf{z}$ are produced by the Graph WaveNet and retain information about the past.

Instead of guessing the density directly, we transform white noise
$\boldsymbol{\varepsilon}\sim\mathcal N(\mathbf 0,\mathbf I)$ through a sequence of $R$ invertible transformations (Fig.~\ref{fig_gnf})
\begin{equation}
    \boldsymbol{\varepsilon}=\boldsymbol{\xi}^{(0)}
    \xrightarrow{\mathcal T_1}
    \boldsymbol{\xi}^{(1)}
    \xrightarrow{\mathcal T_{2}}
    \;\cdots\;
    \xrightarrow{\mathcal T_{R}}
    \boldsymbol{\xi}^{(R)}=\boldsymbol{\xi}(t). \label{eq:flow_chain}
\end{equation}

According to the rule of change-of-variables, the conditional density is
\begin{equation}
    p\bigl(\boldsymbol{\xi}\,\big|\,\mathbf{c}\bigr)=
    p(\boldsymbol{\varepsilon})\;
    \prod_{r=1}^{R} \;
    \Bigl|\det\Bigl(
        \partial\mathcal T_r(\boldsymbol{\xi}_{r-1}, \mathbf{c})/\partial\boldsymbol{\xi}_{r-1}
     \Bigr)
    \Bigr|^{-1}.
\end{equation}
Therefore, learning the noise reduces to constructing analytically invertible transformations $\mathcal T_r$ with tractable Jacobian.

Here we adopt the architecture of Generative Flow with Invertible 1x1 Convolutions (GLOW)~\cite{kingma2018glow}. Each transformation $\mathcal T_r[\mathbf c]$ repeats three reversible operations
in a fixed order.  

First, an \emph{ActNorm} operation rescales and shifts the noises of every cell 
\begin{equation}
    \boldsymbol{\xi} (n_i^\comp, t) \longmapsto \exp\bigl[\mathbf{s}(n_i^\comp, t)\bigr] \circ \boldsymbol{\xi}(n_i^\comp, t) + \mathbf{t}(n_i^\comp, t)
\end{equation}
where $\mathbf{s}(n_i^\comp, t)$ and $\mathbf{t}(n_i^\comp, t)$ are functions of $\mathbf{c}(n_i^\comp, t)$ modeled by a MLP. It allows the fluctuations of each cell to have its own amplitude and baseline, which are determined by the conditional embedding of its local environment $\mathbf c(n_i^{\comp},t)$.

Second, an invertible $1\times1$ convolution linearly mixes the channels \emph{inside} each cell,
\begin{equation}
    \boldsymbol{\xi} (n_i^\comp, t) \longmapsto  \mathbf{W}(n_i^\comp, t) \, \boldsymbol{\xi}(n_i^\comp, t)
\end{equation}
where $\mathbf{W}(n_i^\comp, t) = \text{MLP}\bigl[\mathbf{c}(n_i^\comp, t)\bigr]$ is an invertible matrix with $\det\mathbf{W}\neq0$. It captures the correlation between different components of the noise vector $\boldsymbol{\xi}(n_i^\comp, t)$ for each cell, by allowing extra affine transformations such as rotation and stretch in addition to scaling by ActNorm. 

Finally, correlations between neighboring cells are introduced by a \emph{graph-coupling} step. 
In this step, the noise vector of each cell is $\boldsymbol{\xi}=(\boldsymbol{\xi}_1,\boldsymbol{\xi}_2)$. A graph convolution is run on the current cell graph
$\mathcal G_{\comp}(t)$, taking $\boldsymbol{\xi}_1$ and $\mathbf c$ as inputs:
\begin{equation}
    \mathbf{s},\mathbf{t}=
    \text{GCN}_{\mathrm{couple}}
    \bigl(\mathcal G_{\comp}(t), \boldsymbol{\xi}_1,\mathbf{c}\bigr).
    \label{eq:gcn_st}
\end{equation}
It yields the scale $\mathbf{s}$ and shift $\mathbf{t}$ fields to further act on the other half of the noise vector:
\begin{equation}
    \boldsymbol{\xi}_2\; \longmapsto \;
    e^{\mathbf{s}} \circ \boldsymbol{\xi}_2 +\mathbf{t},
    \qquad
    \boldsymbol{\xi}_1\;\text{left unchanged}.
    \label{eq:coupling_update}
\end{equation}
Since only $\boldsymbol{\xi}_2$ is modified, the Jacobian is triangular and its
determinant simply reduces to the products of the entries of $\mathbf{s}$, while the graph convolution allows the coupling of the noise vectors among neighboring cells.

Repeating this three-step sequence allows subtle, long-range correlations to emerge while determinant remains easily tractable, so the likelihood of any observed fluctuation can be evaluated analytically. All these transformations are continuously modulated by the conditional embeddings $\mathbf{c}_t$. 
Consequently, the learned noise distribution adapts to changes in neighborhood geometry, biochemical state, as well as mechanical deformation, providing a conditionally normalized description of the fluctuations.

\subsection{Experimental methods}
The experimental datasets employed here are obtained from previously published work \cite{Devany2021}. In brief, MDCK cells were cultured in DMEM supplemented with 10\% FBS and 2mM L-Glutamine. Cells were plated on polymerized collagen gels and allowed to form a confluent monolayer overnight. The monolayer was imaged with a Nikon spinning disk microscope system with a 37C 5\% CO2 incubator using a 20x multi-immersion objective. To obtain images of membranes and nuclei, cells were treated with lentivirus to stably express CACNG2-Halotag and p27-ck-snaptag. Cells were treated for 1 hour with JF-646 halotag ligand and TMR-snaptag ligand, washed once with PBS and imaged in normal growth medium. Cells were imaged at 10 minute intervals.

\subsection{Data pre-processing}
MDCK data were analyzed using previously published methods \cite{Devany2021}. A segmentation algorithm using the phase  stretch transform method \cite{Asghari_Jalali_2015} was used to obtain outlines of the cells and nuclei. Standard functions in Matlab were used to obtain cell and nuclear areas, aspect ratios, centroids, and perimeters. Nuclei centroids were tracked over time using the Simpletracker algorithm \cite{Tinevez_Cao}. Nuclear trajectories with cell parameters were used as inputs to the GNN. Additional data without nuclei labeled were obtained directly from the publication \cite{Devany2021} and analyzed using the same algorithm but using the cell centroid for tracking.

Fly wing data were obtained from a previous publication \cite{Etournay2015}. Segmented data were obtained and analyzed with tracking algorithm described for MDCK. 

MDCK data with cell cycle information were obtained from a previous publication \cite{Devany2023}. For each cell the average intensity of each FUCCI marker within the boundary was determined and based on a threshold the cell was assigned to a cell cycle state. Cells were tracked using the cell centroid and cell cycle state was added as a node parameter for the GNN.

ERK signaling data were obtained from a previous publication \cite{Boocock2020, Hino2020}. Nuclei with ERK reporter signal were segmented using Ilastik \cite{Berg2019}. The average ERK ratio within each segmented nucleus, the nuclear area, aspect ratio, perimeter and centroid were obtained in Matlab. Simpletracker was used to track these properties for each cell over time as above.

\end{document}